\documentclass[prl,twocolumn,showpacs,preprintnumbers,amsmath,amssymb]{revtex4}

\usepackage{graphicx}
\usepackage{dcolumn}
\usepackage{bm}

\begin{document}

\title{Empirical Proton-Neutron Interactions and Nuclear Density Functional
Theory:~Global, Regional and Local Comparisons}

\author{M.~Stoitsov$^{1-3}$, R.B.~Cakirli$^{4,5}$,  R.F.~Casten$^{4}$,
W.~Nazarewicz$^{1,2,6}$, W. Satu{\l}a$^6$} \affiliation{$^1$Department of Physics
and Astronomy, University of Tennessee, Knoxville, TN 37996, USA}
\affiliation{$^2$Oak Ridge National Laboratory, Oak Ridge, TN
37831, USA} \affiliation{$^3$Institute of Nuclear Research and
Nuclear Energy, Bulgarian Academy of Sciences, Sofia, Bulgaria}
\affiliation{$^4$Wright Nuclear Structure Laboratory, Yale
University, New Haven, CT 06520, USA} \affiliation{$^5$Department
of Physics, University of Istanbul, Istanbul, Turkey}
\affiliation{$^6$Institute of Theoretical Physics, Warsaw
University, Warsaw, Poland}
\date{\today}

\begin{abstract}
Calculations of nuclear masses, using nuclear density functional theory,
are presented for even-even nuclei spanning the nuclear chart. The
resulting binding energy differences can be interpreted in terms of
valence proton-neutron interactions.  These are compared globally,
regionally, and locally with empirical values. Overall, excellent
agreement is obtained. Discrepancies highlight neglected degrees of
freedom and can point to improved density functionals.
\end{abstract}

\pacs{21.30.Fe, 21.10.Dr, 21.60.Jz, 71.15.Mb}
\maketitle

As with other many-body systems, the structure of the atomic
nucleus depends on the interactions of its constituents, protons
and neutrons. These interactions, reflecting the strong and
Coulomb forces, and the Pauli Principle, are complex.
Nevertheless, their understanding is critical to interpreting
nuclear structure and its evolution with $N$ and $Z$. Similar
issues arise in other finite complex systems, such as nanostructures,  
and there is
increasing overlap in the theoretical tools applied. In nuclei,
where two kinds of fermions come into play,
the proton-neutron (p-n) interaction plays the key role in the
development of long-range collective correlations, including
non-spherical shapes. Due to the shell structure of
nuclei, p-n interactions of the valence (open shell) nucleons are
the most important.

Since nuclear masses embody the sum of all nucleonic interactions,
they provide a laboratory in which it is possible to isolate and
extract specific interactions using appropriate mass differences
\cite{jensen84}. In particular, the average interaction of the
last two protons with the last two neutrons in an even-even
nucleus  is given by the following double difference of binding
energies \cite{1,2}:
\begin{equation}\label{filter}
\begin{array}{rl}
\delta V_{pn} {(Z, N)}=& \displaystyle \frac{1}{4}\left[
\left\{B(Z,N)-B(Z,N-2)\right\}  \right. \\ & \\
-& \displaystyle \left. \left\{B(Z-2,N)-B(Z-2,N-2)\right\} \right]
\end{array}
\end{equation}
With the 2003 mass evaluation \cite{4}, it became possible to
evaluate a much larger set of $\delta{V}_{pn}$ values. These have
revealed \cite{5,6,7,8} striking bifurcations near closed shells
and systematic patterns spanning major shells \cite{5}; a
correlation between $\delta{V}_{pn}$ values and growth rates of
collectivity \cite{6}; and intriguing patterns in specific regions
\cite{7}. While simple calculations with schematic zero-range
interactions give reasonable results in the deformed rare earth
nuclei, they fail completely in the actinides \cite{7}. Clearly, a
more sophisticated approach is needed.

The indicator (\ref{filter}) involves  masses of four neighboring
even-even nuclei. Theoretical understanding of the behavior of
$\delta V_{pn}$ throughout the whole nuclear chart thus calls for
an approach that is capable of predicting nuclear masses with
arbitrary $Z$, $N$ values. Such an approach must fulfill several
strict requirements. First, it should be rooted in microscopic
theory. Second, it must be general enough to be confidently
applied to regions of the nuclear landscape whose properties are
largely unknown. Third, it should be capable of handling
symmetry-breaking effects resulting in a variety of intrinsic
nuclear deformations. These requirements are met by density
functional theory (DFT) in the formulation of Kohn and Sham
\cite{ks65}. The main ingredient of the nuclear DFT
\cite{Ben03,kluge} is the energy density functional describing
conditions locally around each nucleon. This can be realized by
expressing the functional in terms of local nucleonic densities
and currents. The energy functional is augmented by the pairing
term describing nuclear superfluidity \cite{Perl}. When not
corrected by additional phenomenological terms, standard
functionals, treated self-consistently, reproduce total binding
energies with an \textit{rms} error of 1.5 to 4 MeV
\cite{Ben03,Lunney,Ber05}. However, they have been successfully tested
over the whole nuclear chart for a broad range of phenomena, and
usually perform better when applied to {\it energy differences}
and other global nuclear properties such as radii and nuclear
deformations.

As pointed out in Ref.~\cite{3}, from Eq.~(\ref{filter}), $\delta
V_{pn}$ approximates the mixed partial derivative
\begin{equation}\label{part}
\delta V_{pn} (Z, N) \approx \frac{\partial^2 B}{\partial Z
\partial N}.
\end{equation}
For nuclei with an appreciable neutron excess ($T_z$$>$1), the average
value $\widetilde{\delta V}_{pn}$ probes the symmetry energy term in the
macroscopic mass formula \cite{rafalski,leptodermous}:
\begin{equation}\label{average}
\widetilde{\delta{V}}_{pn}
{\approx} 2 \left(a_{\rm sym}+ a_{\rm ssym}A^{-1/3}\right)/A,
\end{equation}
where $a_{\rm sym}$ and $a_{\rm ssym}$ are the symmetry and
surface-symmetry  energy coefficients, respectively. For nuclei
with ${N}\sim{Z}$, $\delta{V}_{pn}$ also contains the Wigner
energy which can be extracted by taking differences between values
of $\delta V_{pn}$  in neighboring nuclei~\cite{3}. On top of
$\widetilde{\delta V}_{pn}$, detailed fluctuations of $\delta
V_{pn}$ carry important information about shell effects and
many-body correlations.

It is the purpose of this Letter to present the first results of
large-scale microscopic calculations for $\delta{V}_{pn}$ within
the DFT framework. Our work demonstrates that, since
$\delta{V}_{pn}$ is a relative quantity, calculations with
realistic functionals can reproduce empirical values
(\ref{filter}) to significantly better than 100 keV in many mass
regions. This is of the same magnitude as local fluctuations in
empirical $\delta{V}_{pn}$ values. Hence, these calculations can
assess the adequacy of currently used energy density functionals
and help search for improvements in such aspects as their density
dependence and dynamical correlation effects.

The present large-scale  calculations of nuclear masses are based
on the HFB+THO code described in Ref.~\cite{Sto03}. Our DFT
approach  is based on a self-consistent solution of the
Hartree-Fock-Bogoliubov  equations with an approximate
Lipkin-Nogami treatment of pairing followed by an exact particle
number projection. Calculations are performed in 
a transformed harmonic
basis spanning 20 major shells. In the particle-hole channel we
employed the SkP \cite{Dob84} and SLy4 \cite{Cha98} Skyrme
functionals. The pairing functional corresponds to a
density-dependent $\delta$ interaction of Ref.~\cite{pairing}.
Unless otherwise indicated, a mixed-type pairing was employed.
Overall, over 1000 nuclei were studied. As $\delta V_{pn}$
corresponds to a second derivative, in order to obtain reliable
values, calculations required very high numerical accuracy. In
fact, we found out that $\delta V_{pn}$ provides an excellent
check on the precision of calculations. (A similar conclusion was
drawn earlier \cite{8} in the context of the accuracy of measured
masses.)

We first present a global comparison of experimental and
theoretical $\delta{V}_{pn}$ values in Fig.~1. To accommodate the
wide range of $\delta{V}_{pn}$ values, while preserving the
visibility of the microstructure for heavier nuclei, the panels
use different vertical scales. The overall trends in the data are
well reproduced by Eq.~(\ref{average}) with the SkP values of
$a_{\rm sym}$=30\,MeV and $a_{\rm ssym}$=--45 MeV
\cite{leptodermous}. This confirms empirically the importance of
the surface symmetry term. Superposed on this secular decrease
there are considerable fluctuations. The most dramatic effect is
the empirical singularities in light $N$=$Z$ nuclei. This is well
understood \cite{2,3} as resulting in part from  $T$=0 (p-n)
pairing interactions which are not considered in our DFT model.
The regional fluctuations on the right seem to have substantial
differences between theory and experiment, but a more detailed
view provides a more accurate perspective on regions of agreement
and regions where the calculations are missing key ingredients.
\begin{figure}[ht]
\includegraphics[width=0.48\textwidth]{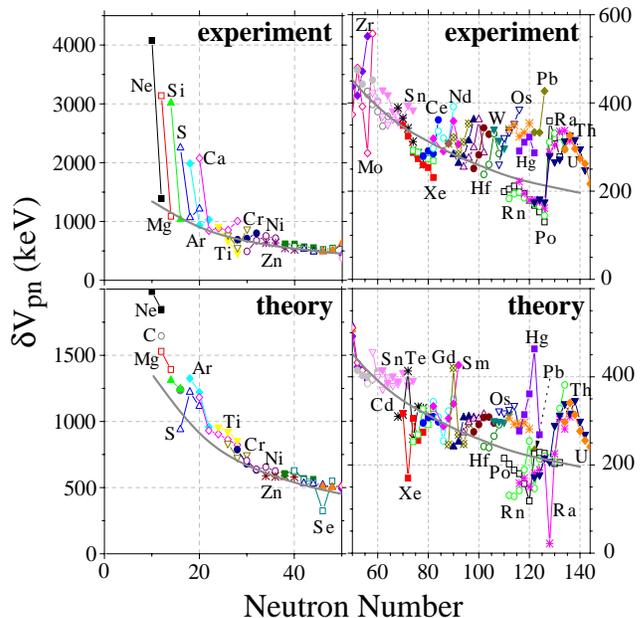}
\caption{Comparison of empirical $\delta{V}_{pn}$ values (top)
across the nuclear chart with DFT calculations using the SkP
functional (bottom). The plot includes stable nuclei as well as
available results for the trans-Pb nuclei from Po to U. The
average $\delta{V}_{pn}$ values from Eq.~(\ref{average}) are shown
by a solid gray curve.}
\end{figure}

To pursue this, Fig.~2 shows four isotope chains comprising
vibrational systems (Cd), two deformed chains (Ra, U), and proton
magic Pb nuclei.
\begin{figure}[htb]
\centerline{\includegraphics[width=0.45\textwidth]{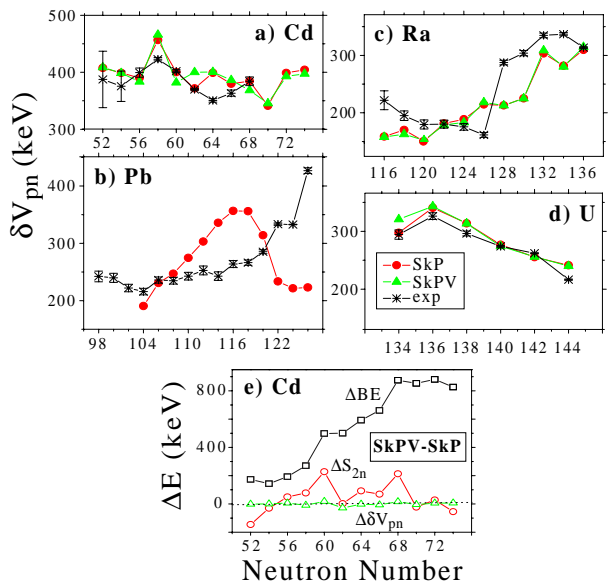}}
\caption{Panels (a-d)
display empirical and calculated $\delta{V}_{pn}$ values for Cd,
Pb, Ra, and U isotopes. Calculations are based on the SkP
functional with mixed (SkP) or volume (SkPV) pairing. Panel (e)
shows differences in binding energies (BE), 2-nucleon separation
energies ($S_{2n}$) and $\delta{V}_{pn}$ values obtained in SkPV
and SkP for the Cd isotopes (see text).}
\end{figure}
The calculations were carried out with the SkP functional with the
standard mixed pairing term (SkP) or with the volume pairing
(SkPV). For Cd and U, the agreement with experiment is remarkably
good for both functionals, and quite acceptable for Ra, generally
within a few to 10's of keV. Given the average mass accuracy of
several MeV, this demonstrates a striking ability to probe
specific interactions by exploiting the filtering capabilities of
mass difference indicators. The Ra-U panels show opposite
empirical trends, both well reproduced by the calculations. Yet,
there are discrepancies in Ra for $N$=128-134 which point to
missing physics. Here, an octupole, reflection asymmetric, degree
of freedom, not included in the present calculations, plays an
important role \cite{octupoles}. The Pb nuclei show large
discrepancies, reflecting the inadequacy of the current DFT
approach for describing strong dynamical changes in ``core''
structure near magic numbers.

Another interesting feature seen in the SkP and SkPV results of
Fig.~2 is that the choice of pairing only weakly affects $\delta
V_{pn}$. This is because, at least for nuclei away from the
$N$=$Z$ line, the pairing correlation energy can roughly be
written as a sum of independent proton (p-p) and neutron (n-n)
contributions; hence, the leading components of pairing are
expected to be filtered out by the indicator (\ref{part}). It is
instructive to inspect the effectiveness of this filter. Figure~2e
shows the differences for the SkP and SkPV calculations in masses,
$S_{2n}$, and $\delta{V}_{pn}$. Although the mass differences
($BE$) range from up to nearly an MeV, the first derivatives,
$S_{2n}$, are much closer (typically differing by $<$100 keV),
and, for the mixed derivatives $\delta{V}_{pn}$, the differences
are essentially zero.

\begin{figure}
\includegraphics[width=0.40\textwidth]{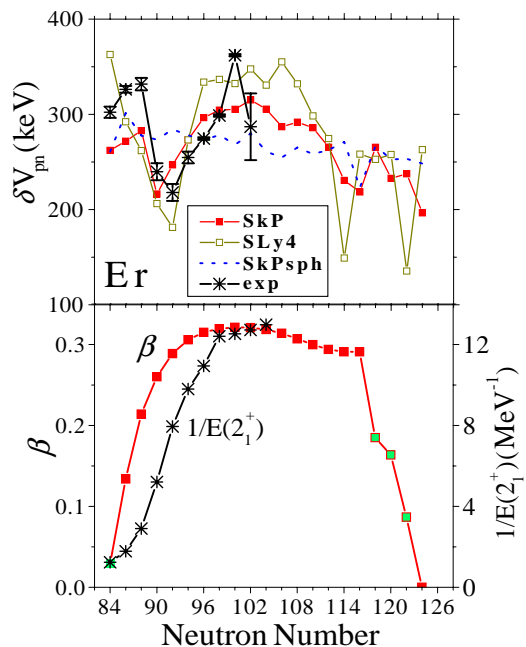}
\caption{Top:~$\delta{V}_{pn}$ values for Er isotopes.
Calculations are  performed with SkP and SLy4 functionals with
mixed pairing. The SkP results constrained to a spherical shape
(SkPsph) are marked by a dotted line. Bottom:~Calculated absolute
values of the quadrupole deformation $\beta$ (red/green squares
for prolate/oblate shapes) together with experimental values of
1/$E(2^+_1)$.}
\end{figure}
Figure 3 (top) shows three sets of calculations for Er spanning a
spherical-deformed transition region. Generally, the SkP and SLy4
functionals (both with the same mixed  pairing) reproduce the
trends in the data where empirical $\delta{V}_{pn}$ values are
known ($N$=84-102), including larger values for the spherical
nuclei, a sharp drop in the transition region (${N} \sim$ 90-94),
and an increase beyond that. The lower part shows the calculated
expectation values of $\beta$. These are compared to empirical
values of 1/$E$(2$^+_1$) which is a useful measure of shape
correlated with the moment of inertia. The calculations reproduce
the transition region quite well, including the onset of
deformation near $N$=90,92 (perhaps slightly shifted to lower
neutron numbers), and its saturation above $N$=100. Beyond
$N$=102, where no data exist, the SkP and SLy4 results differ
markedly, and in a way rather typical for many cases. The SLy4
results often show much larger fluctuations than the more robust
SkP results, which also agree better with the data than other
interactions tested. Figure 3 (top) also shows SkP calculations
where the shape was \textit{constrained} to a spherical shape.
These calculations, of course, are not suitable for transitional
or deformed nuclei such as Er and their neighbors, but serve here
as a pedagogical benchmark. Excursions above the spherical
reference (around the middle of the shell) and below this line
(\textit{e.g.}, near $N$=90) exhibit the effects of shell
structure on quadrupole correlations. In this context, it is
interesting to recall that the microscopic origin of quadrupole
deformations in nuclei can, within the nuclear DFT, be attributed
to p-n interactions \cite{defpn}.

\begin{figure}
\centerline{\includegraphics[width=0.35\textwidth]{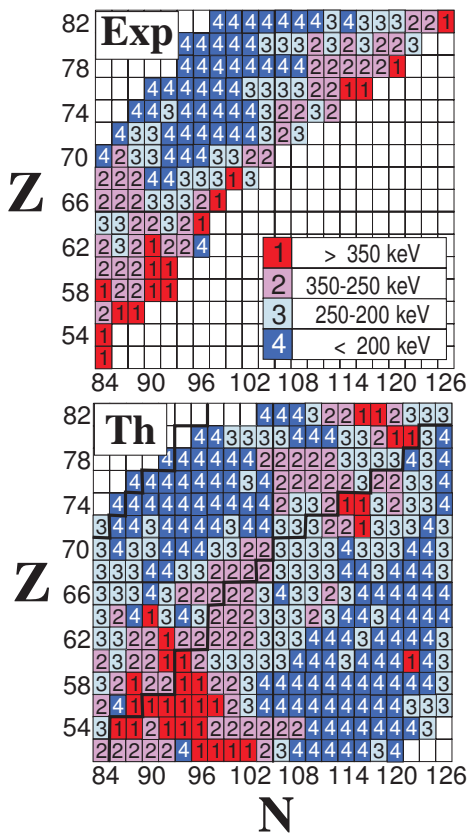}}
\caption{Empirical and
calculated $\delta{V}_{pn}$ values for the major shells $Z$=50-82,
$N$=82-126, color coded according by magnitude [red for the
largest values, blue for the smallest]. (Upper panel taken from
Ref.~\cite{5}). The zig-zag lines in the theoretical panel enclose
nuclei with known empirical $\delta{V}_{pn}$ values. Four boxes
within these lines are empty since some of the nuclei involved are
calculated to lie beyond the proton drip line and cannot be
calculated reliably by the current DFT model.}
\end{figure}
The global and local comparisons in Figs.~1-3 are complemented by
a regional perspective in Fig.~4. Arguments \cite{5} from generic
shell structure can account for the overall systematics. Normal
parity orbits fill high-$j$, low-$n$ orbits early in a major shell,
but low $j$-high $n$ orbits at the end. Thus one expects the
largest $\delta{V}_{pn}$ values near the diagonal, where there is
similar fractional filling of proton and neutron shells and
therefore maximum spatial overlap. At the time of Ref.~\cite{5},
there were no realistic calculations that could be brought to bear
on this empirical phenomenon. However, the power of the nuclear
DFT approach is shown by the present calculations which reproduce
very well \textit{both} the general magnitude of $\delta{V}_{pn}$
values and their variations across a shell, including larger
values near the diagonal. Figure 4 illustrates another point.
Overlap arguments again suggest \cite{6} larger empirical
$\delta{V}_{pn}$ values where protons and neutrons are filling
similar quadrants (lower left and upper right) and smaller values
in dissimilar regions (upper left). It was speculated \cite{6}
that $\delta{V}_{pn}$ values would also be small in the currently
data-free lower right quadrant. The calculations strikingly
confirm this qualitative idea and enhance the need for new mass
measurements in this region.

To conclude, we have presented the first large-scale  microscopic
DFT calculations of $\delta{V}_{pn}$ spanning the nuclear chart,
comparing the results for different functionals, including
different treatments of pairing, and confronting these
calculations with empirical results. Overall, the agreement is
impressive. This is especially significant since the average error
of the calculated binding energies is several MeV. Yet,
proton-neutron interaction energies obtained from binding energy
differences generally match the data to well under 100 keV and,
often, significantly better than that. The agreement is best in
the deformed regions where the mean-field theory is expected to
capture essential physics. This level of agreement can be
exploited to extrapolate to unknown nuclei when three of the four
masses needed for a $\delta{V}_{pn}$ value are known. For example,
using the predicted $\delta{V}_{pn}$ value for $^{238}$U, and the
known masses for $^{238,236}$U and $^{234}$Th, gives a predicted
mass excess of -46325 keV for $^{236}$Th. It would be interesting
to test this compared to the extrapolation from systematics
\cite{4} of -46454 keV. This approach gives useful predictions for
($Z, N-2$) and ($Z-2, N$) nuclei in proton and neutron-rich
regions, respectively. Finally, deviations of $\delta{V}_{pn}$
values do occur, and point to needed improvements in the density
functionals, where specific effects or degrees of freedom enter.
Example were noted at magic numbers, $N$=$Z$ nuclei, and in the
octupole-correlated Ra region.

We are grateful to D.S.~Brenner, and Y.~Oktem for
useful discussions. Work supported by US DOE grant numbers
DE-FG02-96ER40963, DE-AC05-00OR22725, and DE-FG02-91ER-40609. This
research used resources of the National Center for Computational
Sciences at ORNL.

\end{document}